\documentclass[aps,prd,amsmath,amssymb,superscriptaddress,showkeys,showpacs,twocolumn,floatfix]{revtex4-2}
\usepackage{graphicx}
\usepackage[utf8]{inputenc} 
\usepackage{epstopdf}
\usepackage{amssymb}
\usepackage{amsmath}
\usepackage{ulem}
\usepackage{hyperref}
\usepackage[all]{hypcap}% jump to the figure, not to the figure caption

 \newcommand{\ltot}{l}
 
\usepackage{xcolor}

\begin{document}

\title{The cosmic QCD transition for large lepton flavour asymmetries}

%\subtitle{Do we have a subtitle?}


\author{
Mandy M. Middeldorf-Wygas}
\email{e-mail: m.wygas@physik.uni-bielefeld.de}
\affiliation{Fakult\"at f\"ur Physik, Universit\"at Bielefeld, Postfach 100131, 33501 Bielefeld, Germany}

\author{Isabel M. Oldengott} 
\email{e-mail: isabel.oldengott@uv.es}
\affiliation{Departament de Fisica Te\`orica and IFIC, CSIC-Universitat de Val\`encia, 46100 Burjassot, Spain}
 
\author{Dietrich B\"odeker}
\email{e-mail: bodeker@physik.uni-bielefeld.de}
\affiliation{Fakult\"at f\"ur Physik, Universit\"at Bielefeld, Postfach 100131, 33501 Bielefeld, Germany}

\author{Dominik J. Schwarz}
\email{e-mail: dschwarz@physik.uni-bielefeld.de}
\affiliation{Fakult\"at f\"ur Physik, Universit\"at Bielefeld, Postfach 100131, 33501 Bielefeld, Germany}

\date{\today}

\begin{abstract}
We study the impact of large lepton flavour asymmetries on the cosmic QCD transition. Scenarios of \textit{unequal} lepton flavour asymmetries 
are observationally almost unconstrained and therefore open up a whole new parameter space for the cosmic QCD 
transition. We find that for large asymmetries the formation of a Bose-Einstein condensate of pions can occur and identify the corresponding parameter space. In the vicinity of the QCD transition scale, we express the pressure in terms of a Taylor expansion with respect to the complete set of chemical potentials. The Taylor coefficients rely on input
from lattice QCD calculations from the literature. The domain of applicability of this method is discussed.

%\keywords{early Universe \and cosmic QCD transition \and lepton asymmetry}
\end{abstract}

\maketitle

\section{Introduction}
\label{sec:Intro}
The recent direct detection of gravitational waves (GWs) by the LIGO/Virgo 
collaboration \cite{Abbott:2016blz} has revived the interest in phase transitions in the early Universe 
\cite{Caprini:2018mtu}. 
In general, first-order phase transitions can be accompanied by processes that lead to the emission of GWs, while crossovers 
do not lead to a strong enhancement over the primordial GW spectrum. 
Within the Standard Model of particle physics (SM) at vanishing
chemical potentials
both the electroweak transition at $T_{\mathrm{ew}}\sim 160$  GeV as well as the transition of quantum 
chromodynamics (QCD) at $T_\mathrm{QCD}$ $\sim 150$~MeV 
are expected to be crossovers. 
However, many extensions of the scalar sector of the 
Standard Model can give rise to a first-order elelectroweak phase 
transition~\cite{Espinosa:1993bs}. 
Even the simplest extension,
i.e.\  by a real singlet, 
is difficult to exclude at the LHC (see, e.g.~\cite{Cepeda:2019klc}).
There are however much fewer mechanisms known  
that could provide a first-order QCD transition (e.g.\ \cite{Iso:2017uuu,Bodeker:2021mcj,Hambye:2018qjv}), which is why this possibility has obtained considerably less attention.   
The softening of the equation of state during the QCD transition,  be it first order or a crossover, leads to an enhancement of the production of primordial black holes  \cite{Jedamzik:1996mr,Jedamzik:1999am,Byrnes:2018clq}, one of the prime candidates for dark matter, see e.g. \cite{Carr:2020xqk}.

Our current understanding of the QCD phase diagram is as follows. Whether strongly interacting particles exist in the form of quarks and gluons or in the form of hadrons (i.e.\ confined quarks and gluons) depends in general on the temperature $T$ and the baryon chemical potential $\mu_\mathrm{B}$ (and all other potentials associated with relevant conserved quantum numbers) of the system. It is known from lattice QCD \cite{Aoki:2006we,Bhattacharya:2014ara,Ding:2015ona} that at vanishing chemical potentials the transition between these two phases is a crossover, where a pseudocritical temperature is calculated to be $T_{\text{QCD}}=156.5\pm 1.5$ MeV by \cite{Bazavov:2018mes} and 
 $T_{\text{QCD}} = 158.0\pm 0.6$ MeV by \cite{Borsanyi:2020fev}. 
At large baryon chemical potential $ \mu  _ B $
and vanishing temperatures,
in contrast, effective models of QCD (like the Nambu Jona-Lasinio model) predict a
first-order chiral transition \cite{ Asakawa:1989bq}.
This leads to the speculation that there exists a critical line in the 
$(\mu_{B},T)$-diagram which separates the two phases by a first-order transition and which is supposed to end in a second-order critical end point. Functional QCD methods predict this critical end point (CEP) to be located around $(\mu_{B}^{\mathrm{CEP}},T^{\mathrm{CEP}})=(672,93)$ MeV \cite{Gao:2020qsj}.

The standard trajectory of the Universe in the QCD diagram is based on the 
assumption of tiny matter-antimatter asymmetries and is expected to start at large temperatures and low chemical potentials. 
Due to the expansion of space the temperature decreases, and the 
baryon chemical potential %$ \mu_{B}$ 
is expected to remain small roughly until pion annihilation at 
$T\approx m_{\pi}/3$, 
when $\mu_{B}$ starts to approach the nucleon mass. Therefore, as mentioned above, the transition is expected to be a crossover. 
This idea of the standard cosmic trajectory is based on the observation 
that the Universe has an extremely small and well measured baryon 
asymmetry,  
\begin{equation}
b \equiv n _ B /s = (8.70 \pm 0.06) \times 10^{-11},
\end{equation} 
as inferred from \cite{Planck:2018vyg}.
Here $ n _ B  $ is the baryon density, i.e. the number of baryons minus 
the number of antibaryons per volume, and $ s $ is the entropy density. Furthermore, it is reasonable to assume an electric charge neutral Universe, i.e. a vanishing electric charge asymmetry, $q \equiv n_{Q}/s =0$ \cite{Caprini:2003gz}. 
If the lepton flavour asymmetries $l_{\alpha}=n_{L_{\alpha}}/s$ are of the same order of magnitude as $b$, this indeed implies a small
baryon chemical potential $\mu_{B}$ 
at $T \gtrsim m_{\pi}/3$ \cite{Schwarz:2009ii,Wygas:2018otj}. 
However, observational constraints on the lepton asymmetry are much weaker (see e.g.\ \cite{Oldengott:2017tzj}), allowing in principle a lepton asymmetry that is many orders of magnitude larger than the baryon asymmetry. The impact of lepton flavour 
asymmetries on the abundance of weakly interacting dark matter has been studied in \cite{Stuke:2011wz}.

As shown in \cite{Schwarz:2009ii,Wygas:2018otj}, for increasing values of the  lepton asymmetry the cosmic trajectory is shifted 
towards larger values of $\mu_{{B}}$. 
Refs.~\cite{Schwarz:2009ii,Wygas:2018otj} 
assumed the lepton flavour asymmetries $l_{\alpha}$ to have the same values (i.e. $l_{\alpha}=l/3$ 
where $l$ is the total asymmetry). 

In this work, we drop the assumption of equal lepton flavor 
asymmetries that we made in 
Ref.~\cite{Wygas:2018otj}. 
We show that scenarios of unequal lepton flavour asymmetries are even less 
constrained and therefore can have an even larger 
impact on the cosmic trajectory. 
It turns out that the chemical potentials can become so big that they may
affect the nature of the cosmic QCD transition.

This article is organized as follows. In section \ref{sec:Method} we summarize our method to calculate the cosmic trajectory. Section \ref{sec:Leptonasymmetry} discusses the situation of unequal lepton flavour asymmetries. Our main results are presented in section \ref{sec:CosmicTrajectory}, where we also discuss the limitations of our current method. 
Further details are provided in two appendices. We conclude in section \ref{sec:Conclusions}. All expressions of this work are provided in natural units, 
i.e.,~$c = \hbar = k_\mathrm{B} = 1$.

\section{Method}
\label{sec:Method}
In this section, we summarize the method of \cite{Wygas:2018otj} and also reveal an improvement to it in the high temperature regime. For more details we refer the reader to  \cite{Wygas:2019tsx,Wygas:2018otj,Schwarz:2009ii}. 

We assume that 
baryon number, lepton flavour and electric charge,
as well as 
the entropy are conserved in a comoving volume between temperatures
well above the QCD transition,%
\footnote{  In the Standard Model 
baryon number and lepton flavour numbers are conserved
below the electroweak sphaleron freeze-out temperature
$ T_ { \rm sph }  \simeq 131 $ GeV.} 
i.e., $ T\sim 500 $~MeV
and the onset of neutrino oscillations at $T_{\mathrm{}}\sim 10$ MeV.
Each conserved charge $ C \in \{B, Q, L _ \alpha  \} $ 
can be assigned a chemical potential $ \mu  _ C $.
Then from the pressure $ p ( T, \mu  ) $
computed in a grand canonical ensemble 
\begin{align} 
  \exp \left (     p V /T \right ) 
  =  \mbox{tr}\, \exp \left [ 
   \left ( \sum _ C \mu  _ C C    - H \right ) /T \right ]
  \;, 
\end{align}
one can obtain 
the charge densities through
\begin{align} 
  n _ { C } = 
  \frac{ \partial p} { \partial \mu  _ C }  
	\label{nc_1} 
  \;.
\end{align}
For an ideal gas  
this gives 
\begin{equation}
n_C = \sum_i C_i \, n_i (T,\mu_i).
\label{nc_2}
\end{equation} 
where the  sum runs over the contributing individual particle species with 
net number densities $n_i$ and charges $C_i$.
The chemical potentials of the individual particle species $\mu_i$ can be expressed through the chemical potential of the conserved charges $\mu_C$, see e.g. \cite{Kapusta:2006pm,Schwarz:2009ii},
\begin{equation}
\mu_i = \sum_C C_i \mu_C.
\end{equation}
 Depending on the temperature range we either express the charge asymmetries through eq. \eqref{nc_1} or eq. \eqref{nc_2}.

The basic idea of this work and \cite{Wygas:2018otj,Schwarz:2009ii} can be summarized as follows: 
The cosmic trajectory in the QCD phase diagram is determined by the conservation of $l_{\alpha}$, $b$ and $q$, i.e., by the requirement that these quantities 
remain constant throughout the evolution of the Universe 
in the temperature range under consideration.
The  baryon asymmetry can be fixed to 
to its observed value (see above)
and we assume an electric charge neutral Universe, $q=0$ \cite{Caprini:2003gz}. This 
leaves us with the three lepton flavour asymmetries
$l_{\alpha}$ as free input parameters.

In the presence of lepton flavour asymmetries
the  $ \mu_{L_ \alpha} $ are nonzero.
With nonvanishing $ \mu  _ { L _ \alpha  } $ 
the electric charge in the lepton sector
is non-vanishing. 
It has to be compensated by an electric charge in the quark sector.
To obtain total electric charge equal to zero and baryon asymmetry 
(almost) zero, nonzero 
$ \mu  _ Q $ and $ \mu  _ B $ are required.
This implies that lepton flavour asymmetries may have an 
impact on the nature of the cosmic QCD transition if they are sufficiently 
large.

By numerically solving the conservation laws
for $10 \, \mathrm{MeV}<T<500 \, \mathrm{MeV}$ 
for a given set of $l_{\alpha}$ we obtain the Universe's trajectory in six-dimensional $(\mu_{B},\mu_{Q}$, $ \mu_{L_{\alpha}},T)$ space. 
To account for 
the confinement of quarks into hadrons at $T_{\mathrm{QCD}}\sim 150$ MeV
we divide this temperature range into three different regimes that we will describe in the following.
Leptons, on the other hand, can be treated 
at all times as an ideal gas: We calculate their thermodynamic quantities, such as the number densities on the RHS of eq. \eqref{nc_2}, assuming Fermi-Dirac distributions. In the temperature regime of interest it is entirely sufficient to treat neutrinos as massless particles.

\paragraph{Quark-gluon plasma (QGP, $T \gtrsim T_{\mathrm{QCD}}$)}

\begin{figure}
\includegraphics[width=0.5\textwidth]{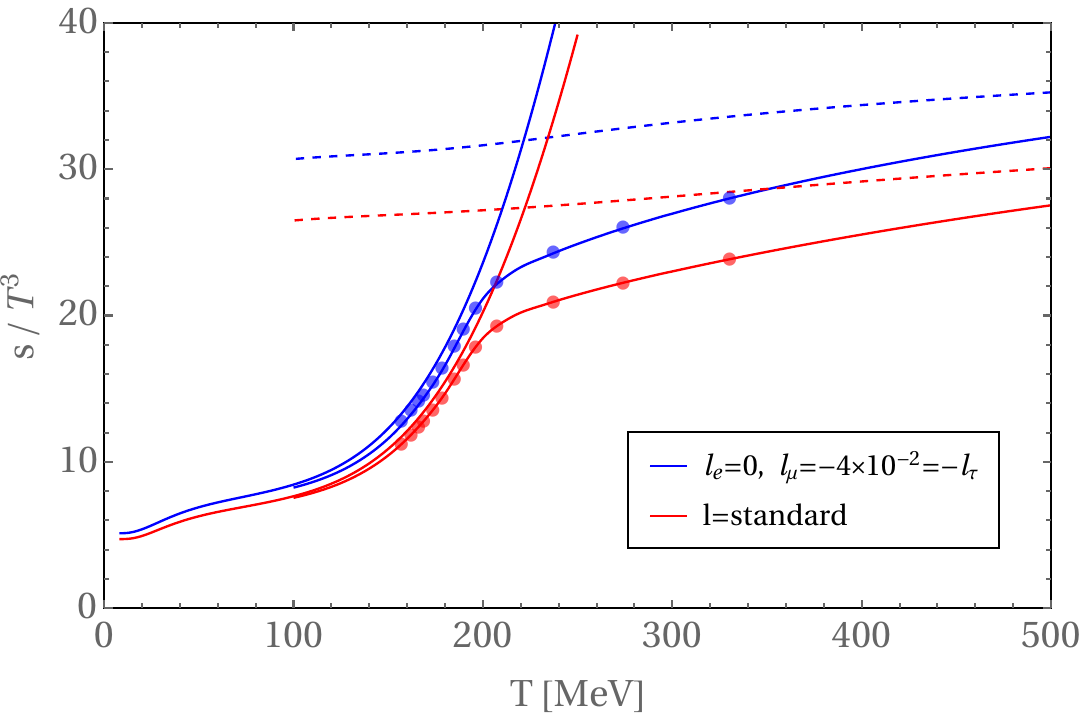}
\caption{Temperature evolution of the entropy density for the standard lepton asymmetry $l=-\frac{51}{28}b$ where  $l_{\alpha}=\frac{\ltot}{3}$, and for the case of unequilibrated lepton flavour asymmetries with $l_e=0,l_{\mu}=-4\times 10^{-2}=-l_{\tau}$. Continuous lines at low temperatures are results for the HRG. The symbols $\bullet$ indicate results obtained using 2+1+1 flavor lattice-QCD susceptibilities \cite{Bazavov:2014yba,Mukherjee:2015mxc}. The dashed lines are the ideal quark gas results. Continuous lines for high temperatures are the results using $\epsilon(T,0)$, and $p(T,0)$ including strong interaction effects according to \cite{Laine:2006cp,Laine:2015kra} 
and the additional contributions 
for non-zero chemical potentials in 
the ideal-gas approximation.}
\label{fig:entropy}
\end{figure}

In \cite{Wygas:2018otj} we treated quarks and gluons as an ideal gas.
Due to considerably strong gluonic interactions this is however not a 
very good approximation.
As we show in fig. \ref{fig:entropy} for the entropy density, 
the ideal-gas curves significantly deviate from the lattice results.
In this work, we therefore use perturbative QCD corrections from  \cite{Laine:2015kra}
to the entropy density $s(T,0)$, the energy density $\epsilon(T,0)$ and the pressure $p(T,0)$.
The calculations in \cite{Laine:2015kra} were performed at 
zero chemical potentials.
To account for non-zero
chemical potentials  we add $s _ {\rm id} (T,\mu)-s _ {\rm id}(T,0)$ where $s _ {\rm id }(T,\mu)$
is the ideal gas entropy.
While these
extra contributions are negligible for small lepton (flavour) asymmetries
they become sizeable for $|l_{\alpha}|\gtrsim 0.01$, as is apparent from fig. \ref{fig:entropy}. The charge asymmetries are calculated in the ideal-gas limit as in \cite{Wygas:2018otj}. In App.\ 
\ref{app:QGPvsIdealGas} we show the impact on the cosmic trajectory 
of adding perturbative QCD corrections to the ideal 
quark gas result.

\paragraph{QCD phase (QCD, $T \approx T_{\mathrm{QCD}}$)}
At temperatures around $T \sim 150$ MeV quarks confine into hadrons and perturbative QCD can no longer be applied. 
As in \cite{Wygas:2018otj}, we therefore make use of susceptibilities $\chi_{ab}$ from lattice QCD to obtain a Taylor expansion of the QCD pressure, 
\begin{equation}
\begin{aligned}
p^\mathrm{QCD}\!(T,\mu)\! &=\!p^{\mathrm{QCD}}\!(T,0)+\frac{1}{2}\mu_a\chi_{ab}(T)\mu_b +\mathcal{O}(\mu^4)\ \\
& \equiv p^{\mathrm{QCD}}_0(T)  +  p^{\mathrm{QCD}}_2(T,\mu) ,
\end{aligned}
\label{eq:pressureQCD}
\end{equation}
with an implicit sum over $a,b\in \{B,Q\}$. 
Baryon and electric charge densities can then
be expressed in terms of the chemical potentials through eq. \eqref{nc_1}.

In \cite{Wygas:2018otj}, we presented results using two different data sets for the lattice-QCD susceptibilities: 
i) continuum 
extrapolated including $u, d$ and $s$ quarks \cite{Bazavov:2012jq} and ii) not continuum-extrapolated 
but including also the $c$ quark \cite{Bazavov:2014yba,Mukherjee:2015mxc}. 
We showed that the inclusion of the charm quark is essential for connecting
the different temperature regimes, which is why we only use the second data set in this work. 
For the entropy density we use the results for
zero chemical potentials from \cite{Laine:2015kra}, 
which interpolate
between HRG and perturbative QCD \cite{Laine:2006cp},  and add the 
non-zero-$ \mu  $ contribution 
as in eq.\ (\ref{eq:pressureQCD})  \cite{Wygas:2018otj}. 

Note that the entropy density obtained from the Taylor series almost perfectly agrees with the ideal quark gas result in fig. 1 
also at large chemical potentials. 
As we will see in the sec. \ref{sec:CosmicTrajectory}, the use of lattice susceptibilities is nevertheless essential in order to connect the cosmic trajectory between the different temperature regimes.

\paragraph{Hadron resonance gas (HRG, $T \lesssim T_{\mathrm{QCD}}$)}
As in \cite{Wygas:2018otj} at low temperatures, we assume an ideal gas of hadron resonances (i.e. thermal distributions), taking into account hadron resonances up to mass $m_{\mathrm{\Lambda(2350)}}\approx 2350 $ MeV $\sim 15T_{\mathrm{QCD}}$ according to the summary tables in%\cite{PDG:2018}
~\cite{Tanabashi:2018oca}.

\section{Large lepton flavour asymmetries}
\label{sec:Leptonasymmetry}

The baryon asymmetry of the Universe is a tiny and well measured quantity. The origin of this number however cannot be explained within the SM of particle physics and gives rise to an active field of research.
The idea of leptogenesis \cite{Fukugita:1986hr} is to create an initial lepton asymmetry that is
partially converted into the baryon asymmetry by electroweak sphaleron processes. 
Therefore, according to the standard picture, the lepton asymmetry of our Universe would be on the same order of magnitude as the baryon asymmetry (i.e tiny), or more explicitly $l=-\frac{51}{28}b$ \cite{Kolb:1983ni}, where the exact numerical pre-factor depends on the assumed particle content of the Universe before the onset of sphaleron processes. This idea however still awaits experimental evidence 
and there are alternative models predicting large lepton asymmetries \cite{Eijima:2017anv,Ghiglieri:2018wbs,Harvey:1981cu,Affleck:1984fy,Canetti:2012kh,Drewes:2021nqr}. Either way, lepton asymmetry is a key parameter to understand the origin of the matter-antimatter asymmetry of the Universe.

When abandoning the assumption of a negligible lepton asymmetry we are left with three lepton flavour asymmetries 
as free input parameters. In principle, those lepton flavour asymmetries can be initially different in size. At $T\approx 10$ MeV neutrino 
oscillations become efficient which can lead to an equilibration of lepton flavour asymmetries such that finally 
$l_{\alpha} \approx \ltot/3$ \cite{Dolgov:2002ab,Wong:2002fa}. It should be noted that depending on the initial values of the lepton flavour 
asymmetries and the mixing angles, equilibration may be only partial, 
i.e., $l_{\alpha} \neq \ltot/3$ \cite{Pastor:2008ti,Barenboim:2016shh,Johns:2016enc}. However, assuming that $l_{\alpha} = \ltot/3$ 
after the onset of neutrino oscillations allows us to obtain constraints on $\ltot$ from the observation of primordial elements, 
i.e., big bang nucleosynthesis (BBN) \cite{Pitrou:2018cgg}, and the cosmic microwave background (CMB) \cite{Oldengott:2017tzj}. 
In \cite{Wygas:2018otj} \textit{for simplicity} we assumed equal lepton flavour asymmetries, $l_{\alpha} = \ltot/3$. 
By numeically solving the conservation laws
as explained in the previous section, we showed that for increasing values of $|\l |$ the trajectory passes through larger absolute values of the chemical potentials $\mu_i$ ($i=B,Q,L_{\alpha}$). For the maximally allowed value $| \ltot | < 1.2 \times 10^{-2}$ from CMB observations \cite{Oldengott:2017tzj}, the lattice susceptibilities allowed to connect the QGP and HRG phases relatively smoothly (given the expected uncertainties from our approximations and the lattice QCD results).  This result however was based on the assumption of equal lepton flavour asymmetries and in this work we investigate the by far less constrained scenario of unequal lepton flavour asymmetries. In that case, since the total lepton asymmetry is conserved during neutrino oscillations, we are still constrained by $| \ltot | <1.2 \times 10^{-2}$ \cite{Oldengott:2017tzj} but the magnitudes of the individual lepton flavour asymmetries could be much larger than $|\ltot|$ as long as they fulfill $| l_e+l_{\mu}+l_{\tau}| < 1.2 \times 10^{-2}$. It was shown in \cite{Mangano:2011ip} that such scenarios also lead to values of the effective number of relativistic degrees of freedom $N_{\mathrm{eff}}$ that are in 
agreement with BBN and CMB observations, i.e., $N_{\mathrm{eff}}\sim 3$.     

\section{Cosmic trajectory}
\label{sec:CosmicTrajectory}

The fact that the individual lepton flavour asymmetries are essentially unconstrained before the onset of neutrino oscillations opens up a huge parameter space for the cosmic trajectory at the times of the QCD transition. This raises the question if large enough lepton flavour asymmetries could change the nature
of the QCD transition. Before exploring this new parameter space in section \ref{sec:Trajectory} we demonstrate  that our current method does not allow to study arbitrary large lepton flavour asymmetries. 

Despite the observational constraint $| \ltot | = | l_e+l_{\mu}+l_{\tau} | <1.2 \times 10^{-2}$, unequal flavour asymmetries provide a lot of parameter freedom. 
For simplicity, in 
the main part of 
this work we only present the results for the scenario $l_{e}=0, l_{\mu}=-l_{\tau}$. We refer the reader to Appendix \ref{app:DiffCases} for more examples of differently distributed lepton flavour asymmetries.
As a supplement to our previous work \cite{Wygas:2018otj} we also discuss the case of equal flavour asymmetries.

\subsection{Pion condensation}
\label{sec:PionCond}

\begin{figure}
\includegraphics[width=0.5\textwidth]{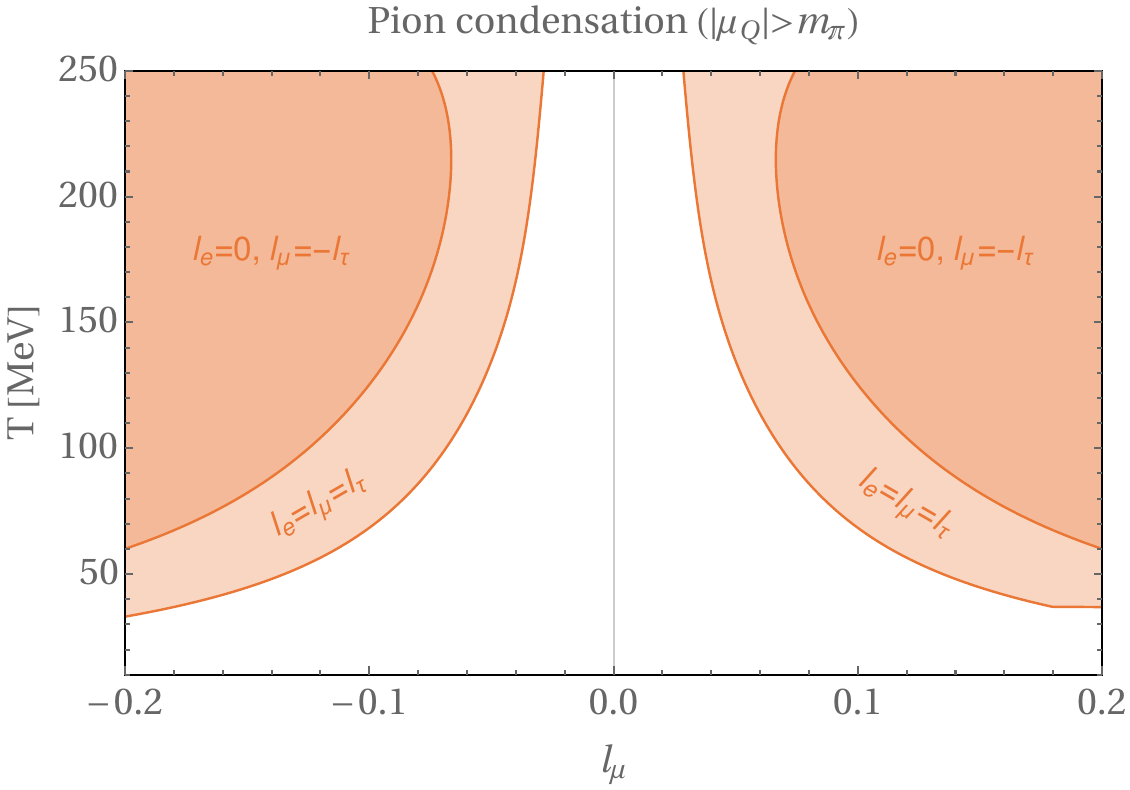}
\caption{Values of the muon lepton asymmetry $l_{\mu}$ for which pion condensation ($|\mu_{Q}|>m_{\pi}$) can occur, depending on the temperature. Dark shaded regions are for the case of unequal lepton flavour asymmetries with $l_e=0,l_{\mu}=-l_{\tau}$, light shaded regions are for equal lepton flavour asymmetries $l_e=l_{\mu}=l_{\tau}=\frac{l}{3}$.}
\label{fig:PionCond}
\end{figure}

Within the HRG approximation, a large chemical potential of the electric charge $\mu_{Q}$ 
can lead to the formation of a Bose-Einstein condensate of pions. 
This happens when the chemical potential of the pion becomes larger than its mass, i.e., $|\mu_{Q}|=\mu_{\pi} \geq m_{\pi}$. While this in general could have interesting consequences such as the formation of pion stars \cite{Abuki:2009hx,Brandt:2018bwq}, our compuation does not apply because we have not included a condensate. 

The condition $|\mu_Q| \geq m_{\pi}$ (and $q=0$) translates non-trivially 
into the $(\mu_B,\mu_{L_e},\mu_{L_{\mu}},\mu_{L_{\tau}})$-$ T $  planes. 
In order to determine the region of parameter space $(\mu_B,\mu_{L_e},\mu_{L_{\mu}},\mu_{L_{\tau}})$ 
in which pion condensation could 
happen, we add $|\mu_Q|=m_{\pi}$ as an additional condition on 
top of the conservation laws. 
While in general the three values for $l_{\alpha}$ are free to choose as input parameters, this extra condition fixes one degree of freedom and therefore only two lepton flavour asymmetries can be chosen freely while the third one is determined at each temperature $T$ by numerically solving the conservation laws \textit{and} $|\mu_Q|=m_{\pi}$.

Fig. \ref{fig:PionCond} shows the solutions of this set of equations for two different parameter choices: equal lepton flavour asymmetries ($\l_e=l_{\mu}=l_{\tau}=\frac{l}{3}$) and unequal lepton flavour asymmetries exemplary for $l_e=0,l_{\mu}=-l_{\tau}$. Note that both of these cases are effectively described by only one degree of freedom and therefore no further input is required. The shaded regions in fig. \ref{fig:PionCond} show for which value of the muon lepton asymmetry $l_{\mu}$ pion condensation may occur at a given temperature. Fig. \ref{fig:PionCond} is consistent with the findings of \cite{Vovchenko:2020crk} which includes a complementary study of the conditions for a pion condensate.
 Since the lepton flavour asymmetries are conserved quantities (before the onset of neutrino oscillations), we should choose the most conservative value for $l_{\mu}$ from fig. \ref{fig:PionCond} in order to avoid the appearance of pion condensation. As evident from fig. \eqref{fig:PionCond}, for unequal lepton flavour asymmetries this constrains the reliability of our method to $|l_{\mu}| \lesssim 0.06$ and for equal flavour asymmetries to $|l_{\mu}|\lesssim 0.03$ (i.e. $|l| \lesssim 0.09$).

\subsection{Applicability of Taylor expansion}
\label{sec:Lattice}

\begin{figure}
\includegraphics[width=0.5\textwidth]{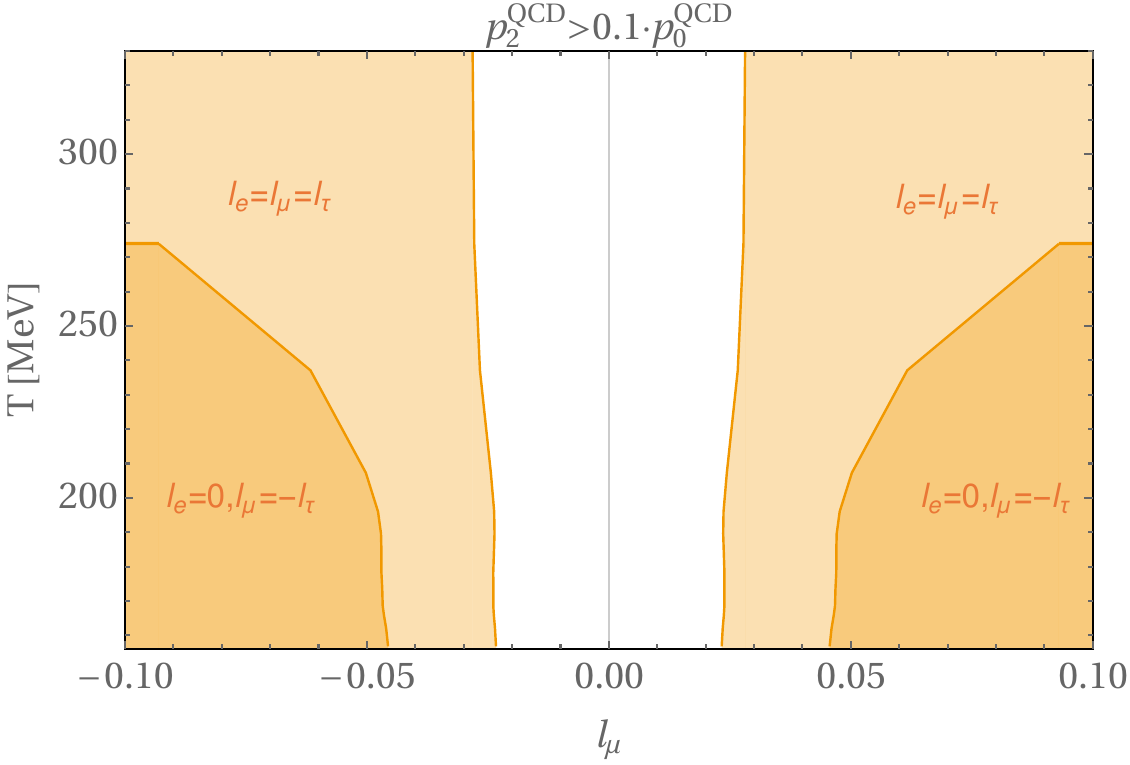}
\caption{Values of the muon lepton flavour asymmetry $l_{\mu}$ for which the use of the Taylor expansion becomes questionable ($p_2^{\mathrm{QCD}}(T,\mu) > 0.1 p_0^{\mathrm{QCD}}(T)$), depending on the temperature. Light shaded regions are for the case of unequal lepton flavour asymmetries, $l_e,l_{\mu}=-l_{\tau}$, dark shaded regions are for equal lepton flavour asymmetries $l_e=l_{\mu}=l_{\tau}=\frac{l}{3}$.}
\label{fig:Taylor}
\end{figure}

Another restriction for our results comes from the truncation of the Taylor expansion  (\ref{eq:pressureQCD}) 
at second order.
It is naturally expected to break down at large chemical 
potentials which restricts the applicability of our 
computation to sufficiently low values of the lepton asymmetries.  
There is however no strict criterion which tells us when exactly 
the use of eq.~(\ref{eq:pressureQCD}) is still justified. 
A reasonable and conservative estimate 
could be given by
\begin{equation}
\begin{aligned}
& p_2^{\mathrm{QCD}}(T,\mu) \leq 0.1 \cdot p_0^{\mathrm{QCD}}(T) \\
\Rightarrow & \frac{1}{2} \mu_a \chi_{ab} \mu_b \leq 0.1 \cdot p_0^{\mathrm{QCD}}(T).
\end{aligned}
\label{eq:convergence_criterion}
\end{equation}

As for the identification of the potential pion condensation region, we extend our numerical code by adding 
eq. \eqref{eq:convergence_criterion} as a sixth condition in addition
to the conservation laws. Again, this reduces the number of 
degrees of freedom by one, such that only two of the three lepton flavour asymmetries are free to chose. 

Similarly to fig. \ref{fig:PionCond}, we show the solution of this set of 
equations for the cases of equal lepton flavour asymmetries
($l_{\alpha}=\frac{l}{3}$) and unequal lepton flavour asymmetries
(again exemplary for $l_e=0,l_{\mu}=-l_{\tau}$) in
fig. \ref{fig:Taylor}.
It turns out that for the unequal case and for the largest temperature value of the lattice data eq. \eqref{eq:convergence_criterion} is never fulfilled such that the orange region in fig. \ref{fig:Taylor} ends below this temperature value. We conclude that the application of the Taylor expansion in eq. \eqref{eq:pressureQCD} is justified for $|l_{\mu}| \lesssim 0.04$ in case of unequal lepton flavour asymmetries and for $|l_{\mu}| \lesssim 0.025$ (i.e. $|l|\lesssim 0.075$) in case of equal flavour asymmetries. These constraints are hence slightly more restrictive than the ones from avoiding pion condensation.

\subsection{Cosmic trajectory for large lepton flavour asymmetries}
\label{sec:Trajectory}

\begin{figure*}
\centering
\includegraphics[width=0.9\textwidth]{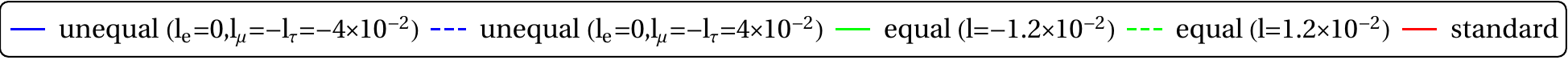}
\includegraphics[width=0.6\textwidth]{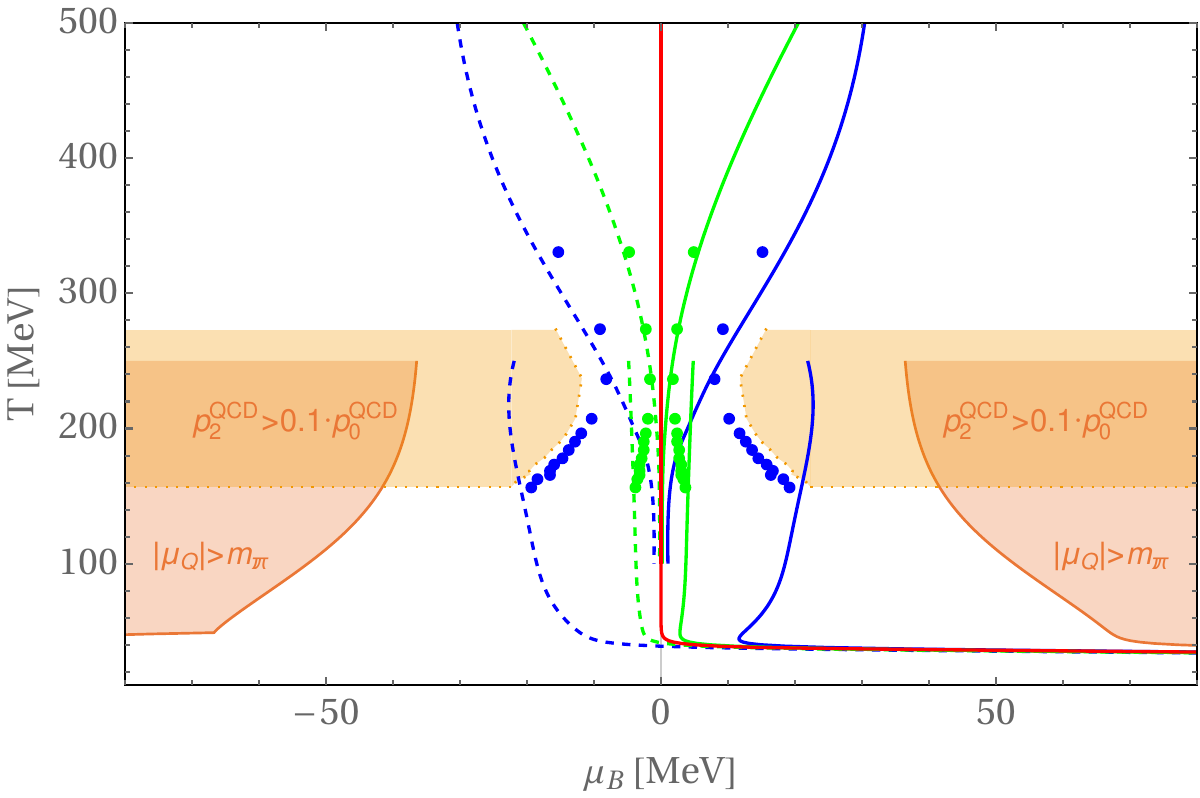}
\includegraphics[width=0.6\textwidth]{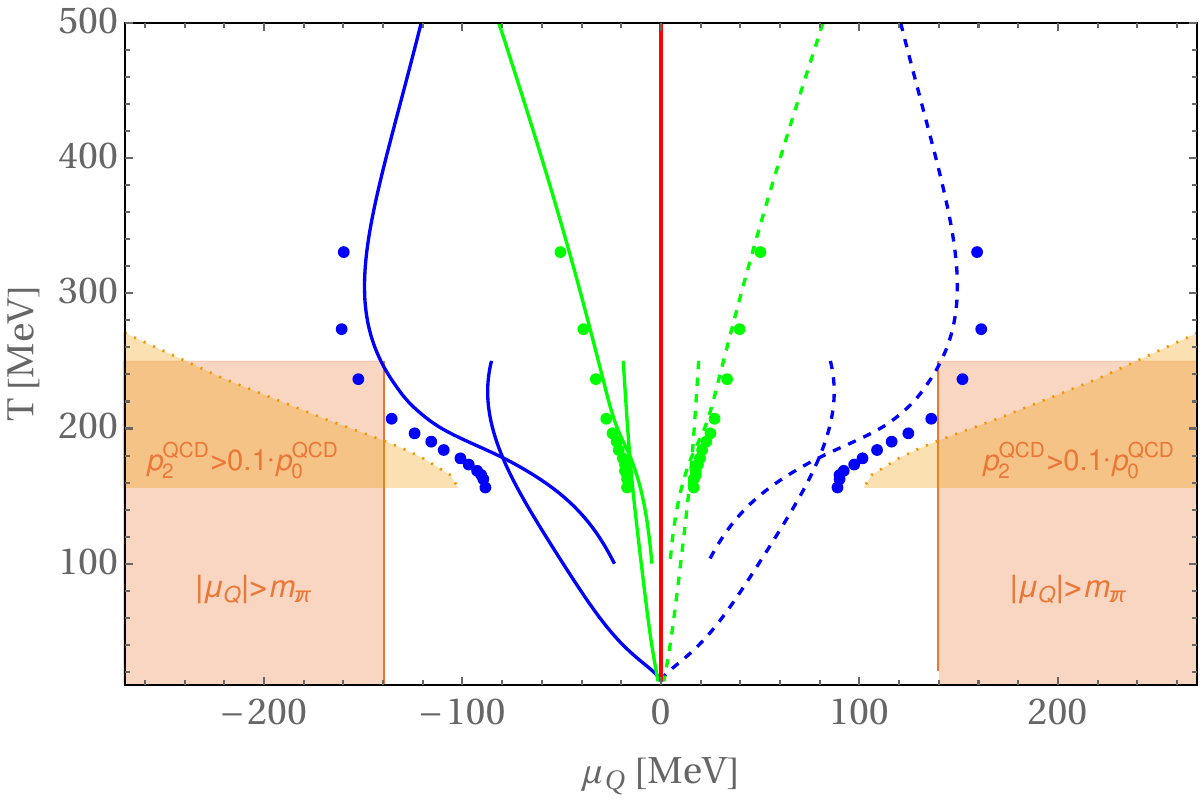}
\caption{Cosmic trajectories projected onto the $(\mu_B,T)$-plane (upper) and the $(\mu_Q,T)$-plane (lower) for different choices of the lepton flavour asymmetries $l_{\alpha}$, calculated for the three temperature regimes described in sec. \ref{sec:Method}. Shaded regions refer to the regions where pion condensation may occur ($|\mu_Q|>m_{\pi}$) and where the applicability of the Taylor expansion becomes unreliable ($p_2^{\mathrm{QCD}}> 0.1 \cdot p_0^{\mathrm{QCD}}$), both discussed in sec. \ref{sec:PionCond} and \ref{sec:Lattice}.}
\label{fig:muB_muQ}
\end{figure*}

In figure \ref{fig:muB_muQ} we show the cosmic trajectory projected on the $(\mu_B,T)$- and $(\mu_Q,T)$-plane for the case of unequal flavour asymmetries ($l_e=0,l_{\mu}=-l_{\tau}$) and the case of equal flavour asymmetries ($l_{\alpha}=\frac{l}{3}$). We present both cases for their maximally allowed values for the lepton (flavour) asymmetries: As we have seen in the previous subsection, the unequal case is restricted by the applicability of the Taylor expansion to $|l_{\mu}| \lesssim 4 \times 10^{-2}$. For the equal case our method is reliable for lepton asymmetries as large as $|l|=7.5\times 10^{-2}$, but  observations of the CMB constrain the lepton asymmetry to $|l|<1.2\times 10^{-2}$ \cite{Oldengott:2017tzj} (i.e. $|l_{\alpha}|<4\times 10^{-3}$). This also implies that all trajectories presented in our previous work \cite{Wygas:2018otj} were neither affected by the restrictions from pion condensation nor by the applicability of
the Taylor expansion. 
For  comparison we also show the standard trajectory (equal lepton flavour asymmtries with $l=-\frac{51}{28}b$). The shaded regions in fig. \ref{fig:muB_muQ} refer to the same regions as figs. \ref{fig:PionCond} and \ref{fig:Taylor} but in the $(\mu_B,T)$- and $(\mu_Q,T)$-planes, i.e. the region where pion condensation occurs and the region where the second order Taylor expansion is not reliable. It turns out that those shaded regions are valid for both cases (equal and unequal lepton flavour asymmetries).  

\begin{figure}
\includegraphics[width=0.5\textwidth]{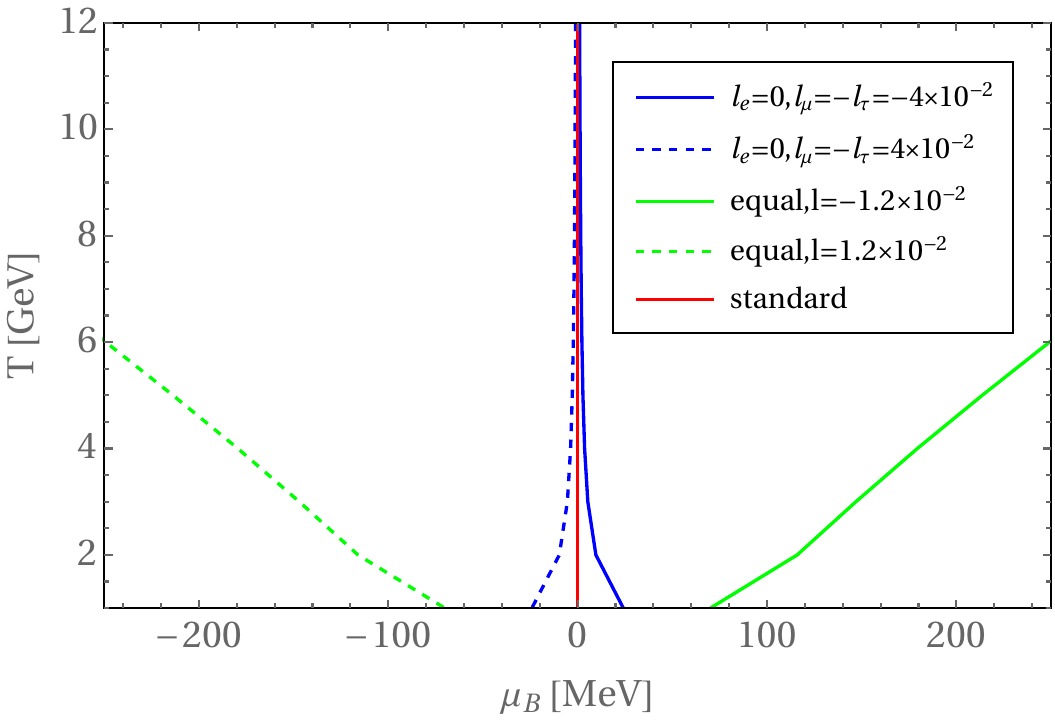}
\caption{Same as upper plot of fig. \ref{fig:muB_muQ} but at higher temperatures.}
\label{fig:highT}
\end{figure}

We see that both cases lead to trajectories reaching sizeable values of $\mu_B$ and $\mu_Q$. Fig. \ref{fig:muB_muQ} furthermore confirms that at the QCD epoch
unequal lepton flavour asymmetries can induce larger chemical potentials than
equal flavour asymmetries,
because the latter case is constrained by CMB observations. 
However, fig. \ref{fig:muB_muQ} indicates that 
at high temperatures the case of equal lepton flavour asymmetries leads to increasing values of $|\mu_B|$ and $|\mu_Q|$ whereas the trajectories of unequal lepton flavour asymmetries seem to bend towards small values of $|\mu_B|$ and $|\mu_Q|$. This is confirmed by fig. \ref{fig:highT} which shows the same cosmic trajectories as fig. \ref{fig:muB_muQ} but for temperatures in the GeV range. In the ultra-relativistic limit the trajectories are indeed only a function of $b$ and the \textit{total lepton asymmetry} $l$ which is equal to zero for the blue curves. This can be shown by a straightforward generalisation of the analytical estimates of $\mu_B$ and $\mu_Q$  in sec. 2.1 of \cite{Schwarz:2009ii}.

As particularly apparent from the $\mu_{Q}$ plot in 
fig. \ref{fig:muB_muQ}, for lepton asymmetries as large as studied in this work, there is also a relatively large gap between the results for the QGP and the QCD phase and as well between the HRG and the QCD phase. We believe that possible reasons for this could be related to the lack of continuum extrapolation and the restricted temperature range of lattice susceptibilities. Another impact could be given by missing finite density effects in the perturbative QCD calculations applied in this work \cite{Laine:2015kra}. 
Furthermore, as stated in sec. \ref{sec:Lattice}, 
there is no strict criterion for estimating
the maximal lepton flavour asymmetries for which the 
use of the Taylor expansion \eqref{eq:pressureQCD} is still justified. 
This also means that there is no  guarantee that the criterion applied in this work, eq. \eqref{eq:convergence_criterion}, is indeed sufficiently conservative.

Note that for the scenario of $l_{\mu}=0, l_e=-l_{\tau}$ the corresponding curves in fig. \ref{fig:muB_muQ} would look extremely similar to the here presented case. The scenario of $l_{\tau}=0, l_e=-l_{\mu}$, in contrast,
is restricted to much smaller values of $\mu_B$ and $\mu_Q$.

\section{Conclusions}
\label{sec:Conclusions}

We extended our previous work \cite{Wygas:2018otj} and studied the cosmic trajectory in the QCD phase diagram for large \textit{unequal lepton flavour asymmetries}. We argued that scenarios of unequal flavour asymmetries are 
much less constrained than the previously studied case of equal lepton flavour asymmetries: 
While for equal lepton flavour asymmetries CMB constraints \cite{Oldengott:2017tzj} restrict the magnitudes of the individual lepton flavour asymmetries to relatively small values, in the case of unequal lepton flavour asymmetries they are essentially unconstrained (as long as their sum fulfills the CMB requirement $| l_{e}+l_{\mu}+l_{\tau} | < 1.2 \times 10^{-2}$ \cite{Oldengott:2017tzj}). Exemplary for the scenario of $l_e=0, l_{\mu}=-l_{\tau}$ we showed that the cosmic trajectory indeed reaches larger $\mu_{B}$ and $\mu_{Q}$ than reachable for equal flavour asymmetries. This extends the parameter space to the region of the QCD diagram in which the nature of the QCD transition is still unknown. In fact, QCD studies until now do not give a conclusive answer about the existence of a CEP and therefore about the possibility of first-order transition. The lack of a full theoretical understanding of the QCD phase diagram actually calls for a 
study of the phenomenological consequences of a first order vs.\ 
cross-over transition at high lepton flavour asymmetries, which perhaps allows us to rule out one of the possibilities. At the same time, the impact of lepton asymmetries on the QCD epoch offers a very interesting perspective to gain insights to the origin of the matter-antimatter asymmetry of the Universe:  If for large enough $l_{\alpha}$ the transition turns out to be first order, the prospect of measuring the GW spectrum with pulsar timing arrays \cite{Tiburzi:2018txc} would offer a way to observationally constrain individual lepton flavour asymmetries (before the onset of neutrino oscillations). 

However, we also showed that our current method is 
not capable to be applied to lepton flavour asymmetries which imply \textit{significantly} larger values of $\mu_B$ and $\mu_Q$ than already studied 
in our previous work \cite{Wygas:2018otj}. 
When the electric charge chemical potential exceeds the pion mass, a Bose-Einstein condensation of pions might form. For both scenarios of equal and unequal asymmetries we determined under which conditions this may happen. While the possible formation of a pion condensate is a phenomenon that must be explored further \footnote{See \cite{Vovchenko:2020crk} for a study of the signal in primordial gravitational waves from a pion condensate.}, in practice for our method it simply implies that our treatment of the low-momentum modes of pions is not sufficient any longer. This could be circumvented by including the possibility of a pion condensate into our method. However, the more serious 
restriction to our method comes from the applicability of a Taylor expansion, in
which we use lattice QCD susceptibilities. We showed that for $|l_{\mu}| \gtrsim 4 \times 10^{-2}$ the chemical potentials become as large that $p_2^{\mathrm{QCD}}(T,\mu) > 0.1 \cdot p_0^{\mathrm{QCD}}(T)$, i.e., the second order contribution becomes a sizeable correction to the zeroth order contribution, which makes the Taylor series approach questionable. This problem could be relaxed by the use of higher order contributions to the QCD pressure; such are however currently not available from lattice QCD calculations including the charm quark. 

Alternatively to the application of lattice susceptibilities might be the application of functional QCD methods \cite{Fischer:2018sdj} which do not encounter any problems in the regime of large chemical potentials and predict the existence of a CEP at $(\mu^{\text{CEP}}_{u/d},T^{\text{CEP}})\approx(200,110)$ MeV \cite{Gao:2020qsj}. The access to the phase structure in the whole $(\mu,T)$-plane could offer a consistent platform to investigate the cosmic trajectory at QCD temperatures, even though the truncations introduced in functional QCD methods would bring in some ambiguities on determining the phase structure quantitatively.

Our final conclusion is that large lepton flavour asymmetries still allow for the possibility of a first-order cosmic QCD transition.  
Extending our study to sufficiently large lepton flavour asymmetries with the presently applied method described in \cite {Wygas:2018otj}, based on 
Taylor expansions around vanishing chemical potentials, is however not possible and further improvements are required. 
 
\begin{acknowledgements}
We thank Fei Gao, Frithjof Karsch, Jürgen Schaffner-Bielich and Christian Schmidt for interesting discussions. We acknowledge support by the Deutsche Forschungsgemeinschaft (DFG) through the Grant No. CRC-TR 211 ``Strong-interaction matter under extreme conditions''. M. M. M.-W. acknowledges the support by Studienstiftung des Deutschen Volkes. I. M. O. acknowledges support from FPA2017-845438  and the Generalitat Valenciana under grant   PROMETEOII/2017/033. 
\end{acknowledgements}

% Change the style file for arxiv version!
%\bibliographystyle{utcaps} 
%\bibliographystyle{spphys}           

\appendix

\section{Impact of perturbative QCD corrections in the high-temperature regime}
\label{app:QGPvsIdealGas}

\begin{figure}
\centering 
\includegraphics[width=0.38\textwidth]{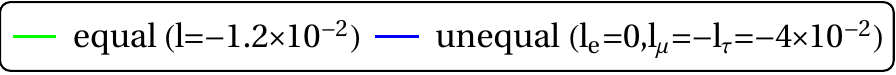}
\includegraphics[width=0.5\textwidth]{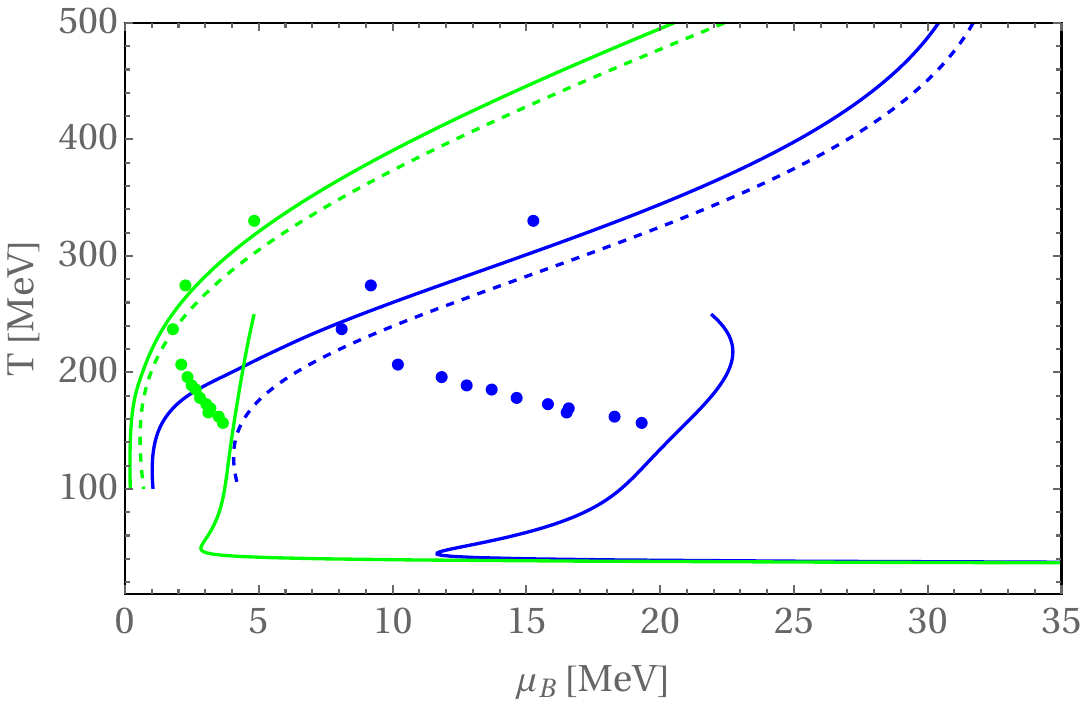}
\includegraphics[width=0.5\textwidth]{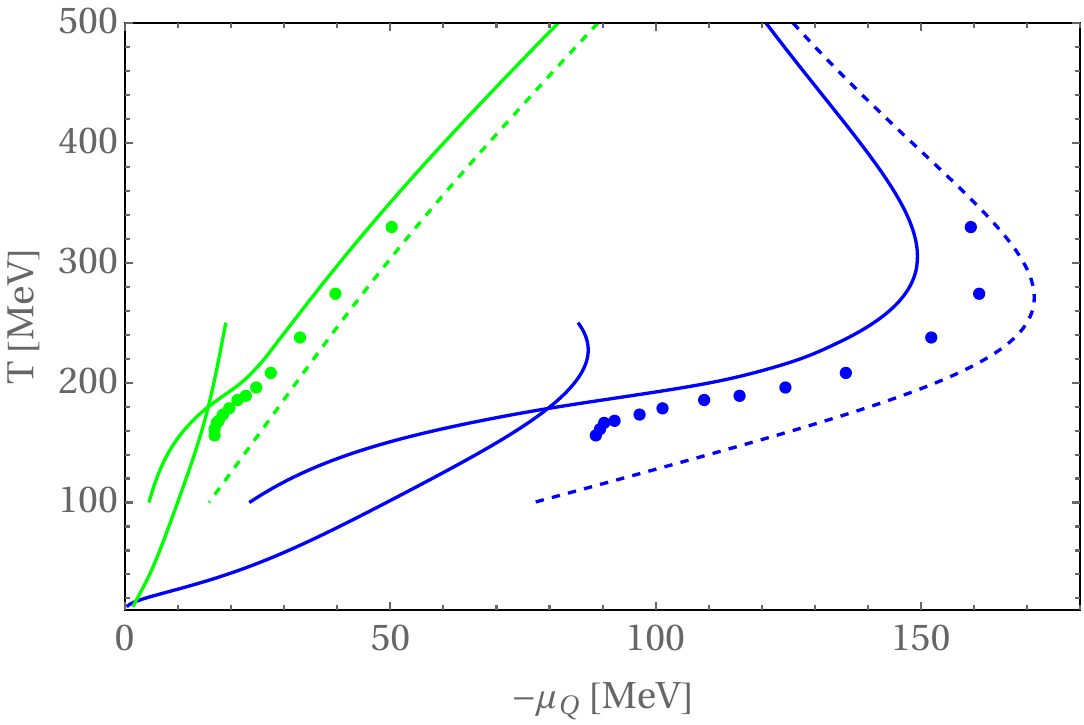}
\caption{Cosmic trajectories projected onto the $(\mu_B,T)$-plane (upper) and the $(\mu_Q,T)$-plane (lower) for different choices of the lepton flavour asymmetries $l_{\alpha}$. Solid lines include corrections from perturbative QCD \cite{Laine:2015kra} (applied in this work), dashed lines assume an ideal gas of gluons and quarks (applied in \cite{Wygas:2018otj}).}
\label{fig:AppendixA}
\end{figure}

As described in the main text of sec. \ref{sec:Method}, in this work we improved our method in the high-temperature regime upon our previous work \cite{Wygas:2018otj}: While we described quarks and gluons
at high temperatures
as an ideal gas in \cite{Wygas:2018otj}, we here take into account corrections from perturbative QCD \cite{Laine:2015kra}. In this appendix, we show how the inclusion of
these corrections impacts the cosmic trajectory. Figure \ref{fig:AppendixA} shows that
the perturbative corrections (solid lines) significantly
shift  the cosmic trajectories compared to the ideal quark gas
(dashed lines) applied in \cite{Wygas:2018otj}.

\section{Different cases of unequal lepton flavour asymmetries}
\label{app:DiffCases}

\begin{figure*}[htp]
\includegraphics[width=0.49\textwidth]{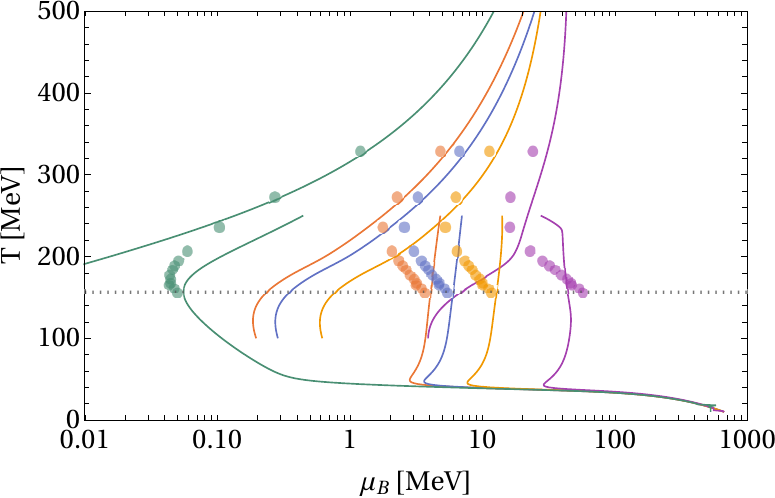}
\hspace{0.1cm}
\includegraphics[width=0.49\textwidth]{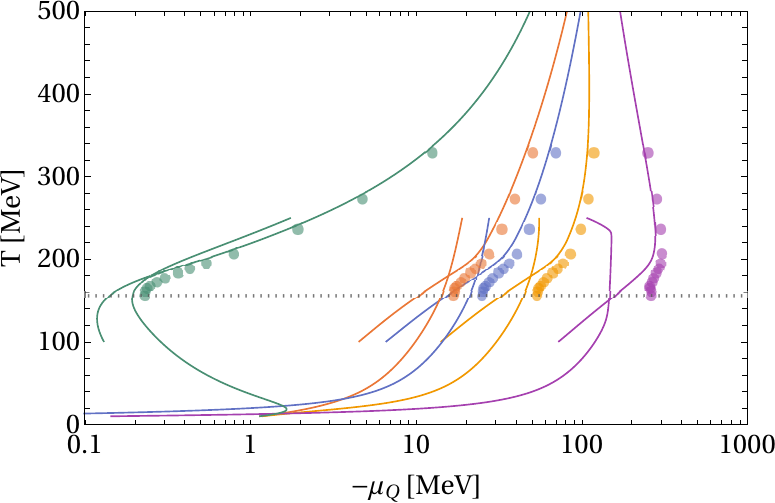}

\vspace{0.5cm}
\includegraphics[width=0.49\textwidth]{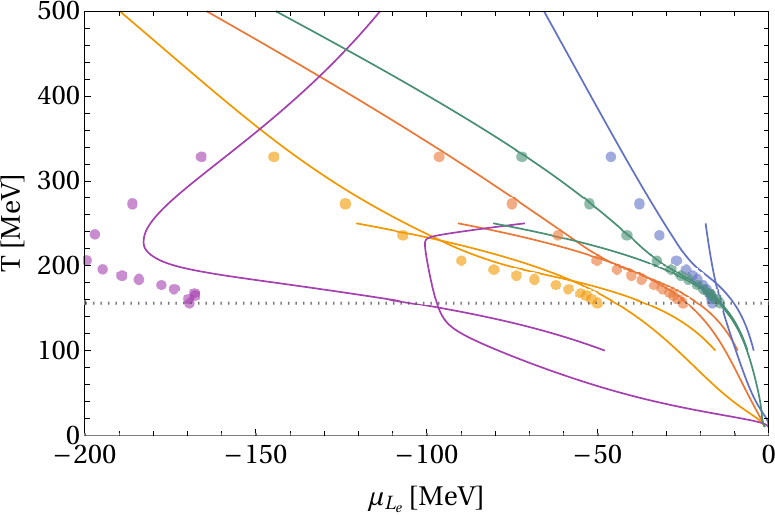}
\hspace{0.1cm}
\includegraphics[width=0.49\textwidth]{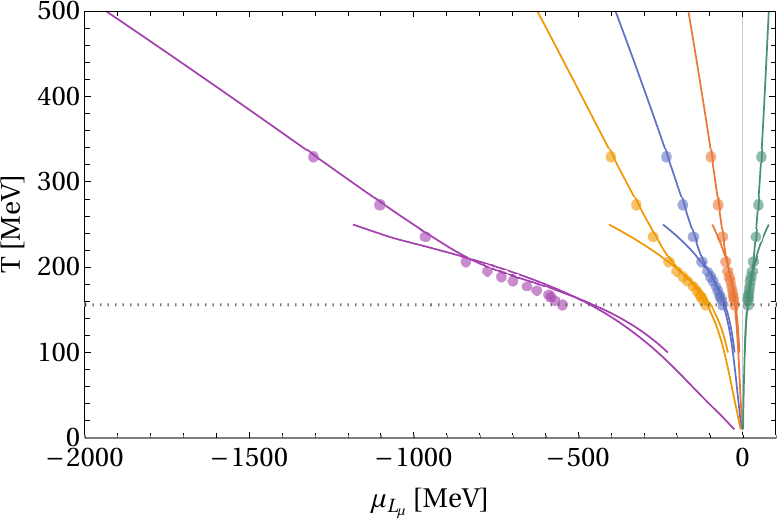}

%\centering
\vspace{0.5cm}
\includegraphics[width=0.49\textwidth]{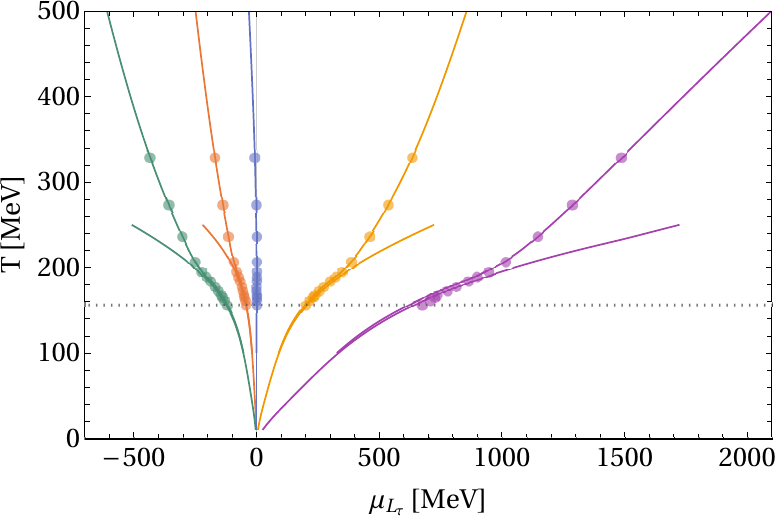}
\hspace{1cm}
\begin{minipage}{0.3\textwidth}
\vspace{-6.3cm}
\includegraphics[width=\textwidth]{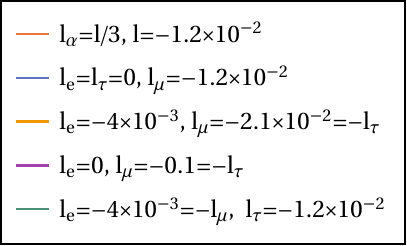}
\end{minipage}
\caption{\label{fig:Mu_noneq} Temperature evolution of conserved charge chemical potentials for different cases of unequal lepton flavor asymmetries. (Top left) Baryon chemical potential $\mu_B$. (Top right) Electric charge chemical potential $-\mu_Q$. (Middle left) Electron lepton flavor chemical potential $\mu_{L_{e}}$. (Middle right) Muon lepton flavor chemical potential $\mu_{L_{\mu}}$. (Bottom left) Tau lepton flavor chemical potential $\mu_{L_{\tau}}$. 
Notations as before.}
\end{figure*}

In this appendix, we show the cosmic trajectories for a variety of different choices of the lepton flavour asymmetries $l_{\alpha}$. Note that a common plot with the restrictions to our method, as in fig. \ref{fig:muB_muQ}, is not feasible since the different trajectories refer to different contours for pion condensation and the applicability of the Taylor expansion. 

\bibliography{Literature}   

\end{document}